# Ferromagnetism in (In,Mn)As alloy thin films grown by metalorganic vapor phase epitaxy

A. J. Blattner and B. W. Wessels[*]

*Department of Materials Science and Engineering and Materials Research Center, Northwestern University, Evanston IL 60208*

**Abstract**

Ferromagnetic properties of $In_{1-x}Mn_xAs$ thin films were investigated. Room temperature ferromagnetic order was observed in nominally single-phase films with $x$ = 0.01-0.10. Magnetization measurements indicated that these $In_{1-x}Mn_xAs$ samples had a Curie temperature of 333 K. The Curie temperature was independent of Mn concentration. The temperature dependent magnetization along with the magnitude of the saturation magnetization and microstructural data indicate that the source of the high-temperature ferromagnetism in single-phase films is not attributable to MnAs nanoprecipitates. The high transition temperature is attributed to the presence of near neighbor Mn pairs.

[*] Corresponding author. Tel.: +1-847-491-3219; fax: +1-847-491-7820;
*E-mail address*: b-wessels@northwestern.edu (B. W. Wessels).

## 1. Introduction

Manganese based III-V diluted magnetic semiconductors (DMSs) such as (In,Mn)As and (Ga,Mn)As continue to receive a great deal of interest due to the observation of ferromagnetism [1]. These materials incorporate a small percentage of magnetic ions within the semiconductor host. For (Ga,Mn)As, Curie temperatures up to 110 K have been reported [2]. However, for many applications ferromagnetic semiconductors with a Curie temperature above room temperature are required. Recently room temperature ferromagnetism has been observed in (Ga,Mn)N [3] exhibiting Curie temperatures up to 370 K. In the case of (In,Mn)As, single-phase films grown by molecular beam epitaxy (MBE) at temperatures less than 280°C were ferromagnetic with Curie temperatures of 35 K or less [4,5]. The low Curie temperature is attributed to the low Mn concentration in these alloys. Films grown at higher temperatures also exhibited ferromagnetism, but with Curie temperatures greater than 300 K and saturation magnetization near 40 emu/cm$^3$ [4]. This observation of room temperature ferromagnetism, however, was attributed to MnAs precipitate formation in the host matrix [6].

We have recently reported the deposition of single-phase (In,Mn)As films at higher growth temperatures using metalorganic vapor phase epitaxy (MOVPE) and have achieved nominally single-phase, epitaxial films as determined by x-ray diffraction for compositions as high as $x =$ 0.14 using growth temperatures greater than 475°C [7]. All (In,Mn)As films exhibited p-type semiconducting behavior with carrier concentrations as high as $2 \times 10^{19}$ cm$^{-3}$ and resistivities between 0.01 and 0.11 Ω-cm. Preliminary variable field Hall measurements exhibited an anomalous Hall effect for an In$_{0.90}$Mn$_{0.10}$As sample [8]. Magnetization measurements indicated that the films were ferromagnetic with Curie temperatures in excess of room temperature [9].

In this article, we report on the magnetic properties of a series of $In_{1-x}Mn_xAs$ films with $x$ = 0.01-0.10. Room temperature ferromagnetism was observed in these samples and $T_c$ = 333 K was measured. The presence of substitutional Mn as dimers and trimers as a possible source of the ferromagnetism with high Curie temperature is discussed.

(In,Mn)As films were prepared using atmospheric pressure metalorganic vapor phase epitaxy at 475-530°C on semi-insulating GaAs(001) substrates. Detailed growth procedures have been previously described [7].

## 2. Structural Characterization

Figure 1(a) is a θ-2θ x-ray diffraction pattern for a 300 nm thick $In_{0.90}Mn_{0.10}As$ film grown at 520°C. Strong $In_{0.90}Mn_{0.10}As$(002) and (004) alloy peaks are observed near 29° and 61°. The GaAs(002) peak and the beginnings of the GaAs(004) peak are apparent near 31° and 65° respectively. No other peaks above the noise of the background were detected. The volume fraction of MnAs precipitates, if present, is estimated to be below a volume fraction of 0.1%. The fraction of manganese incorporated into the nominally single-phase film was determined using energy dispersive x-ray spectroscopy (EDS). Fitting of the EDS spectrum indicated the relative concentration of Mn to In to be 10% for this sample.

MOVPE growth of (In,Mn)As at temperatures less than ~ 475°C results in the formation of (In,Mn)As films with hexagonal MnAs as a second phase. Figure 1(b) is a θ-2θ scan of an $In_{1-x}Mn_xAs$ sample grown at 460°C with Mn/In = 0.10. Additional MnAs(101) and MnAs(102) reflections can be seen near 32.0° and 42.3° respectively.

Plan view scanning transmission electron microscopy (STEM) imaging was also conducted in order to determine the microstructure of these films. Figure 2 is a STEM dark field image of an $In_{0.99}Mn_{0.01}As$ film. Probe size for these measurements was ~ 2 nm giving a spatial resolution of ~ 2.8 nm [10]. Other than inevitable strain contrast due to dislocations, no obvious mass-thickness contrast due to the presence of MnAs precipitates can be observed. Further high resolution and cross-sectional results have also indicated phase purity and will be published elsewhere [11]. Plan view STEM has also been used to image (In,Mn)As samples that exhibit MnAs precipitate formation and provides excellent contrast between the (In,Mn)As matrix and the MnAs precipitates [11].

## 3. Magnetic Properties

The temperature dependent magnetic properties of a series of nominally single-phase $In_{1-x}Mn_xAs$ samples with $x$ = 0.01-0.10 were measured using a superconducting quantum interference device (SQUID) magnetometer with a magnetic field of 10 kOe applied perpendicular to the plane of the film. Measurements were taken at this relatively large applied magnetic field in order to increase the magnetic signal from the film with respect to the large diamagnetic response of the thick GaAs substrate. Measurements have also been taken at 1000 Oe and exhibit similar temperature dependence. After subtraction of the diamagnetic contribution of the GaAs substrate, the resulting magnetization is shown in Fig. 3. These samples display ferromagnetic ordering and an increase in saturation magnetization with increasing Mn concentration. A Curie temperature of 333 K was measured for these samples. This is larger than the $T_c$ = 318 K usually reported for MnAs. As seen in Fig. 3, the Curie temperature was essentially independent of Mn concentration in contrast to films prepared by MBE [2].

The value of the low temperature saturation magnetization of the (In,Mn)As film is given by $M_s = N_{Mn}g\mu_B J_{Mn}$, where $N_{Mn}$ is the nominal concentration of Mn ions, $g$ is the Landeé factor and is equal to 2 for Mn, $\mu_B$ is the Bohr magneton, and $J_{Mn}$ is the spin of Mn. Using $x$ = 0.10, as determined from EDS measurements, the measured $M_s$ corresponds to a value of $\mu$ = 3.6$\mu_B$ and $J_{Mn}$ between 4/2 and 3/2. A more accurate determination of $J$ is difficult due to the uncertainty in the determination of $x$ at low concentrations by EDS as well as nonuniformity of $x$ in the sample. For comparison, theoretical results by Akai based upon KKR-CPA-LDA (Korringa-Kohn-Rostoker coherent-potential and local density approximation) calculations, predict that for $In_{1-x}Mn_xAs$ with $x$ = 0.06, the ground state is ferromagnetic and the local magnetic moment of Mn is 4.2$\mu_B$ [12].

The observation of an anomalously high Curie temperature for a III-V magnetic semiconductor has been previously attributed to the presence of MnAs as a second phase [4]. However, this phase was not detected in our samples by x-ray diffraction measurements as seen in Fig. 1(a) or by STEM as seen in Fig. 2. In addition, high resolution plan view and cross-sectional TEM measurements on a room-temperature ferromagnetic sample have also exhibited no MnAs second phase formation up to a resolution of 2.8 nm [11].

Furthermore, if large clusters of MnAs were present, they should be readily observed in films with magnetization of 60 emu/cm$^3$ by both x-ray diffraction and TEM. If the magnetization observed in our films was a result of MnAs, which has a saturation magnetization of ~520 emu/cm$^3$ (at 10 kOe) for phase pure materials [13], then the measured magnetization corresponds to a volume fraction of MnAs precipitates of ~ 12%.

Magnetization as a function of applied magnetic field was also measured for In$_{0.90}$Mn$_{0.10}$As at several temperatures. Figure 4 displays the results for these measurements. The total magnetization could be described as an algebraic sum of various contributions. The substrate diamagnetic contribution was subtracted from the total magnetization signal. The remaining data consisted of the paramagnetic and ferromagnetic contributions from the film. At 5 K, the saturation magnetization was 62 emu/cm$^3$ with a remanence of 10 emu/cm$^3$ and a coercive field of 400 Oe. When the temperature was increased to 300 K, the saturation magnetization decreased only to 49 emu/cm$^3$.

## 4. Origin of Ferromagnetism

If the films are indeed single-phase, as the XRD, STEM, and magnetization data indicate, the high $T_c$ observed in the present study cannot be explained within the framework of a hole mediated theory of ferromagnetic semiconductors [1]. The theory predicts that the Curie temperature depends on magnetic ion concentration and hole concentration, which is in contrast to present observations. One possible explanation for the difference between theory and experiment is that the Mn in the alloy films in the present study may not be randomly distributed on In sites but is present as atomic scale clusters. Mean field treatments have shown that Mn clustering can significantly enhance $T_c$. This is a result of the localization of spin polarized holes near regions of higher Mn concentration [14,15]. In addition, calculations based upon local spin density approximation (LDSA) have predicted a strong driving force for the formation of magnetic ion dimers and trimers at second nearest-neighbor sites which are ferromagnetic with $\mu = 4\mu_B$ for Mn [16]. In the case of (In,Mn)As, the Mn is located at second nearest neighbor sites at a pair distance of 4.27Å.

In support of a small Mn cluster model, recent calculations using the full potential augmented plane wave (FLAPW) method indicated that Mn pairs on second nearest-neighbor sites in the zinc blende lattice of $In_{30}Mn_2As_{32}$ are strongly ferromagnetic, whereas distant pairs are weakly ferromagnetic and are well described by an RKKY function [17]. Thus if there were a high concentration of second nearest-neighbor Mn pairs (dimers) in the form of $Mn_2As$, strong ferromagnetism should be observed. The concentration of these dimers would be sensitive to their stability and the kinetics of their formation and presumably would differ depending on film growth temperature and technique [18].

From the measured Curie temperature of 333 K for the (In,Mn)As films the exchange interaction can be estimated. We utilize the Monte Carlo simulation results of Diep and Kawamura [19], to relate $T_c$ and $J_1$, where $J_1$ is the exchange constant for second nearest-neighbor magnetic ions. Using $T_c = 0.447|J_1|/k_B$, where $k_B$ is the Boltzmann constant, the measured $T_c$ of 333 K corresponds to $J_1$ of 64 meV, which is in relatively good agreement

with the value of 92 meV calculated by the FLAPW method [17] and 55 meV calculated by LDSA [16].

The observation that $T_c$ is independent of Mn concentration in our (In,Mn)As films is consistent with a cluster-meditated ferromagnetism model [18, 20]. In this case, the measured $T_c$ is dominated by the intra-cluster exchange energy and not the cluster density. This exchange also determines the magnetic moment, $m$, of each cluster. The measured saturation magnetization is then given by $M_s = mN$, where $N$ is the number of dimers. Assuming that Mn dimers predominate the ferromagnetic interaction, changing the Mn concentration in the film will affect $M_s$ but not $T_c$.

## 5. Conclusion

In summary, we have reported the ferromagnetic properties of nominally single-phase $In_{1-x}Mn_xAs$ magnetic semiconductor thin films. Temperature and field dependent magnetization measurements indicated the presence of room temperature ferromagnetism for a series of films with $x = 0.01$-$0.10$. A Curie temperature $T_c = 333$ K was observed for these films that was independent of Mn concentration. The magnitude of the saturation magnetization and the lack of XRD or STEM evidence of MnAs nanoprecipitates indicated that MnAs was not the likely source of the room-temperature ferromagnetism. A possible mechanism for the observed high $T_c$ based upon the preferred formation of Mn dimers at nearest-neighbor cation sites in the thin films was presented.

**Acknowledgements**

The authors wish to acknowledge the assistance of P. L. Prabhumirashi, N. Erdman, and Prof. V. P. Dravid with STEM measurements and B. Watkins with SQUID measurements. We also wish to acknowledge fruitful discussions with Dr. Yu-Jun Zhao and Prof. A. J. Freeman regarding FLAPW calculations. Extensive use was made of the Materials Research Center Facilities at Northwestern University. This work was supported, in part, by the National Science Foundation through the MRSEC program under grant number DMR-0076097 and the Spin Electronics Program under ECS-0224210.

**Figure Captions**

Fig. 1. (a) θ-2θ x-ray diffraction pattern for an $In_{0.90}Mn_{0.10}As/GaAs(001)$ sample grown at 520°C. (b) θ-2θ x-ray diffraction pattern for an $In_{1-x}Mn_xAs/GaAs(001)$ sample grown at 460°C which exhibited MnAs precipitation (Mn/In=0.10).

Fig. 2. STEM dark field plan view image of $In_{0.99}Mn_{0.01}As$. MnAs precipitate formation was not observed. The film had a Curie temperature of 333 K.

Fig. 3. Temperature dependent magnetization at 10 kOe applied magnetic field for a series of nominally single-phase $In_{1-x}Mn_xAs$ films with $x = 0.01$-$0.10$. The magnetic field was applied perpendicular to the plane of the film and the samples were zero-field cooled.

Fig. 4. Applied magnetic field dependent magnetization at 5, 150, and 300 K for $In_{0.90}Mn_{0.10}As$. The inset shows the complete *M-H* hysteresis loop for this sample at 5 K.

Fig. 1—A. J. Blattner and B. W. Wessels

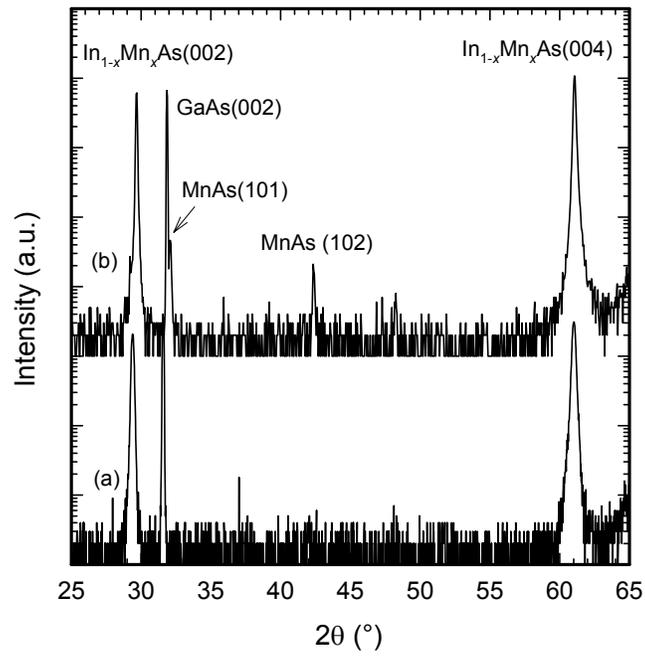

Fig. 2—A. J. Blattner and B. W. Wessels

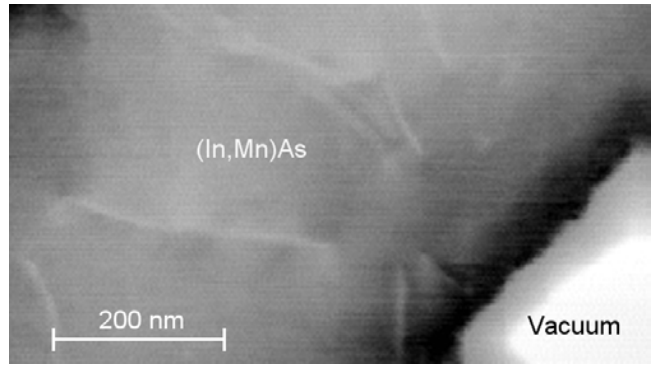

Fig. 3—A. J. Blattner and B. W. Wessels

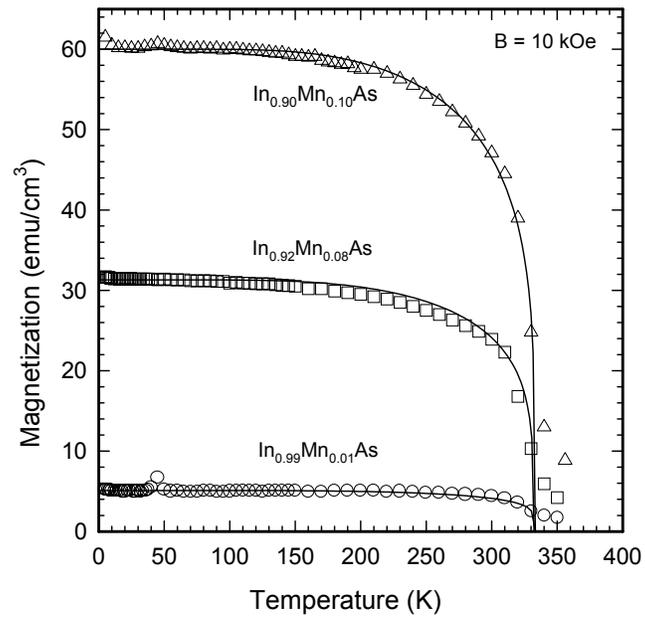

Fig. 4—A. J. Blattner and B. W. Wessels

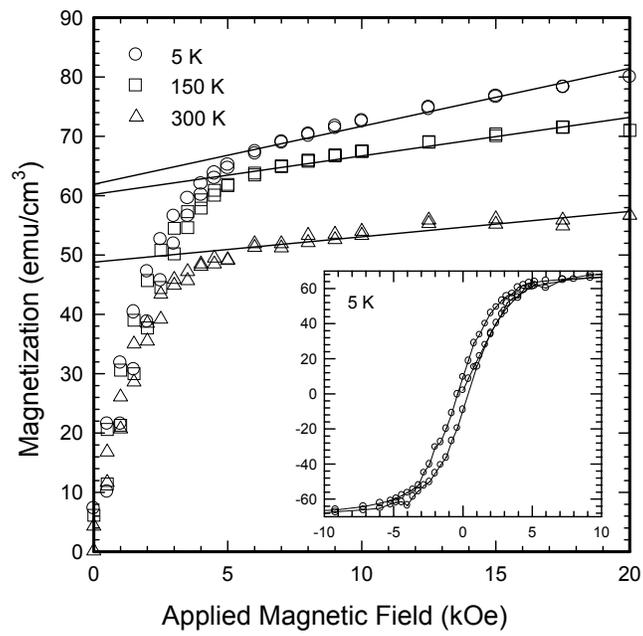